\documentclass{PoS}

\title{Active Galactic Nuclei: present observations and prospects}

\ShortTitle{Active Galactic Nuclei: present observations and prospects}

\author{\speaker{Thierry Courvoisier}\\%
        ISDC Data Centre for Astrophysics\\
        University of Geneva\\
        E-mail: \email{thierry.courvoisier@unige.ch}}

\abstract{INTEGRAL has made a very significant contribution to our understanding of the physics of AGN. We illustrate this progress by  looking  at the considerations that were made at the time of the selection of the INTEGRAL mission and discussing some observations and results obtained in the intervening time.

It is seen that while a number of results pertaining to the variability of AGN have been obtained, the potential of the INTEGRAL data in short timescale variability domain is still largely under exploited. The characteristic variability timescale of AGN, and X-ray binaries, can be understood in the frame of the cascade of shocks model and more generally depends solely on the Compton cooling properties of a hot electron plasma located at a distance proportional to the mass of the object from a soft photon source.

Although INTEGRAL is not a focusing instrument, it is possible to resolve a small but significant fraction of the diffuse extragalactic background at energies higher than 10\,keV.

}

\FullConference{ An INTEGRAL view of the high-energy sky (the first 10 years) - 9th INTEGRAL Workshop and celebration of the 10th anniversary of the launch\\
		 15-19 October 2012\\
		 Bibliotheque Nationale de France, Paris, France}

\begin{document}

\section{Introduction}

When INTEGRAL was designed research on Active Galactic Nuclei (AGN) was an important, but not dominant, element of the scientific goals of the mission. It was known that some AGN emit substantial flux in the hard X-rays and gamma rays, but the link between this emission and the other components of emission was poorly understood. How general this emission is in the different families of AGN was also not clear. It was furthermore expected that the sensitivity of INTEGRAL would be such that long exposures would be needed in order to make significant progress. 

It is interesting to read anew what was written some 20 years ago and see how the science evolved in the mean time and what contribution INTEGRAL did provide.

\subsection{The { \it Red Book} AGN Objectives}

The {\it Red Book} \cite{rb93} was the result of mission studies conducted during phase A that lead to the selection of INTEGRAL as a "medium class" mission of ESA in 1993 and to its complement of instruments.  The {\it Red Book} includes an analysis of the scientific questions to be approached by INTEGRAL and uses them to design an optimum set of instrument properties, taking the available technology into account. In the field of AGN research the {\it Red Book} mentions a number of objectives and questions for the INTEGRAL mission:

\begin{itemize}
\item 
Search for sporadic gamma ray emission from Seyfert galaxies near 500\,keV.
\item
The MeV emission from radio galaxies is discussed, without mention of specific objectives.
\item
Perform detailed studies at low and high spectral resolution in  some 100 AGN.
\item
Locate AGN  with a precision of 1'-5', thus ensuring the "identification" of any newly discovered AGN.
\item
Prevent source confusion, e.g. in the vicinity of 3C\,273. Disentangle the emission from nearby sources.
\item
Detect a considerable number of AGN. Generate logN-logS distributions.
\item
Expect ~100 AGN in the hundreds keV energy range.
\item
Study the electron positron annihilation line for AGN at redshifts larger than 0.001.
\item
Study Compton backscattering features around 200\,keV.
\item
Study absorption and reflection (these words are not used in the text, but this is what is meant in 2012 language).
\item
Study the variability of AGN in the minutes time range.
\item
Measure the polarisation of AGN in detail.
\item
Identify the contribution of AGN to the cosmic diffuse background.
\end{itemize}

\section{Some INTEGRAL AGN results}

The angular resolution of INTEGRAL in the 10s keV range and its sensitivity immediately allowed to lift source confusion, for example in the case of 3C\,273 where previous observations had suggested that the hard X-ray emission in the region might be due to a different source \cite{cetal03}. This notwithstanding, the proportion of "unidentified" sources, i.e. sources for which no counterpart has been measured in other spectral domains, remains close to 1/3   \cite{betal10}. This illustrates that  finding counterparts to weak hard X-ray sources (even when discovered with a non imaging  instrument) remains a non trivial endeavour.

Observations of Seyfert galaxies like NGC\,4151 \cite{letal10} or  Mkn\,509     \cite{petal13} have shown that the emission of these objects is cut-off in the 100\,keV range and that no flux is observed at higher energies. These observations also show that the hard X-ray emission is dominated by  at least one Compton component, mostly attributed to a hot corona surrounding an accretion disk.

Lightcurves of AGN can also be produced with a resolution that is less than one hour. Fig. 1 shows a light curve of 3C\,273. The flux measured in the 1 hour time bins is strong enough to unveil highly significant bin to bin fluctuations by a factor 2 or more, that are common in this section of the lightcurve. The information contained in these lightcurves has yet to be fully exploited. On longer timescales analyses of the hard emission from 3C\,273 by \cite{setal08} have shown some rather unexpected results. One of these is the fact that the variability properties of the flux around 5 keV and in the 20-70\,keV domains are very different, showing that either this emission is made of several components or that the component spanning this energy domain is described by 2 or more parameters (like e.g. a normalisation and a slope) that vary in uncorrelated manners. It was also seen that the 20-70\,keV flux correlates poorly with the UV flux and with the mm flux, while it is closely correlated at zero lag with the V-flux. This shows that the hard X-ray emission in this object is most probably not related to the synchrotron jet, but that it may be the Compton branch of an electron population  emitting synchrotron radiation in the visible domain.

\begin{figure}
\centerline{\includegraphics[angle=0,width=0.7\textwidth]{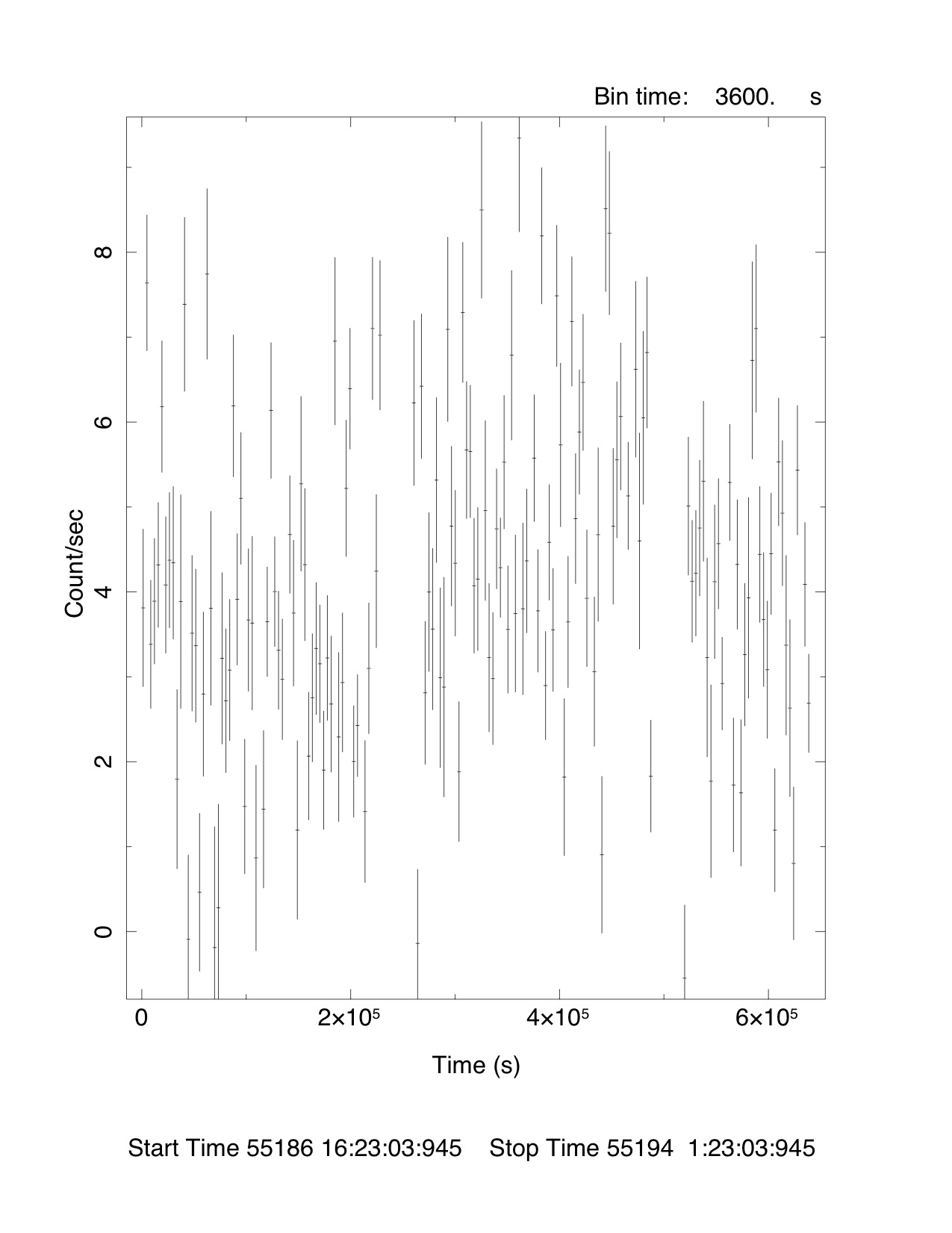}}
%\vspace{-1.2cm}
\caption{\label{fig:3cfield} 20-60\,keV light curve of  3C\,273 with a time resolution of 1 hour (V. Esposito, private communication).
 }
\end{figure}
\section{Cascades of shocks}

Motivated by the long standing difficulties of accounting for the AGN properties with self-consistent disk accretion based models on one side and by the fact that matter accreted from galactic distances losses 999/1000 of its angular momentum and therefore that the accreted material in the central regions of AGN, where the phenomenology takes place, is likely to have a broad angular momentum distribution,  we \cite{candt05} developed a model based on successive shocks in the accretion flow. The implications of this model  for the  hard X-ray emission of AGN and its relationship with the UV emission from these objects were discussed in \cite{iandc09}. It was shown that the X-ray flux is less than the UV flux and that the ratio of X-ray to UV fluxes decreases with increasing object luminosity, a result that had been observed, but which  had remained largely unexplained. 

A further consequence of the cascades of shocks model is the fact that the X-ray time variability $\tau_x$ depends on the mass of the central object $M_{BH}$ and the accretion rate $\dot{M}$ like

\begin{equation}
\tau_x \propto \frac{M_{BH}^2}{\dot{M}} 
\end{equation}

\cite{iandc09b}, a dependence that is very close to that observed by \cite{mchetal06}. It was also subsequently remarked that this result is model independent and  stems solely from the  Compton cooling time of an electron population \cite{iandc12} in a soft photon bath originating in the accretion flow.

\section{INTEGRAL, AGN and the X-ray background}

There are 199 AGN reported in the 5-year AGN catalog of \cite{betal09} and 162 in the catalog of \cite{ketal10}. These (overlapping) populations provide a fair view of the local AGN population properties. It can be seen in these studies that about half of the AGN population is absorbed. A further study of absorption in the AGN population is given in \cite{metal12}. This latter study looked at the X-ray classification of the objects, based on their absorption properties and their optical identification in order to assess the differences that had been reported between those classifications. It results that 88\% of the sources have matching X-ray and optical properties in terms of absorption and Seyfert class, a much better match than previously reported.

The absorption properties of AGN as a function of their luminosity and redshift are an essential tool to understand how they build the X-ray background which, it is reminded, does not match the spectral properties of z=0 AGN populations. In order to progress in this domain, \cite{petal08} analysed all the data available in the field surrounding 3C\,273 and obtained the deepest hard X-ray image to date (see Fig.\,\ref{fig:3cfield}). Individual sources in this image account for some 2.5\% of the diffuse extragalactic X-ray background.

\begin{figure}
\centerline{\includegraphics[angle=0,width=0.8\textwidth]{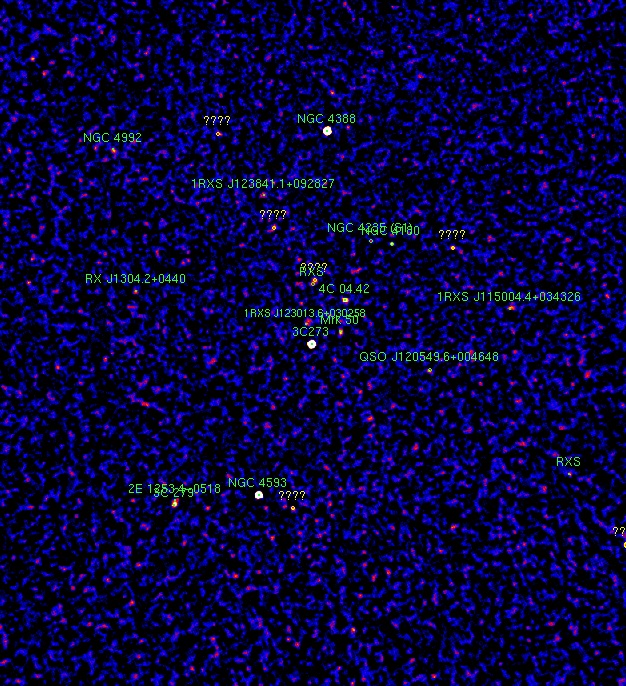}}
%\vspace{-1.2cm}
\caption{\label{fig:3cfield} The field around 3C\,273, the deepest INTEGRAL image. Individual sources represent some 2.5\% of the diffuse X-ray background. The data are discussed in \cite{petal08}.
 }
\end{figure}

A further line of attack on the question of the X-ray spectral shape  in AGN was followed by \cite{retal11}. The unified model implies that the only difference between the different populations of Seyfert galaxies are due to absorption along the line of sight. Studying AGN at energies so high that absorption plays no role anymore is therefore an excellent test of this model. \cite{retal11} showed that, contrary to the expectations derived from the unified model, the properties of Seyfert galaxies differ around 50\,keV. If interpreted in terms of reflection, the difference implies that the reflection component is stronger in Seyfert 2 galaxies than in Seyfert 1 galaxies, implying that the properties of the medium surrounding the central source differ also.  A further result of this analysis is that this increased reflection component is also observed in moderately absorbed sources, thus significantly contributing to resolving the difference between the properties of individual objects and those of the diffuse X-ray background. 

\section{Blazars and Radio Galaxies}

The emission from Blazars is characterised by 2 peaks, the first attributed to synchrotron emission and the second to Compton processes. The first of these peaks is often in the IR-soft X-ray domain, while the second maximum is seen in the very hard radiation. The spectral band in which INTEGRAL is sensitive falls,  for most of the objects, in the valley between the two maxima. It follows that Blazar results are relatively few and are useful mainly to provide some constraints on the low energy part of the Compton emission maximum. It is however also worth noting that \cite{betal12} suggest that IGR\,J12319-0749 might be a z=3.12 Blazar, the highest redshift AGN detected by INTEGRAL (to the knowledge of the author).

A study of the 511\,keV annihilation line was performed in the data from the radio galaxy Cen\,A by \cite{setal05}. Only upper limits were reported.

\section{How have we fared?}

Twenty years have elapsed since the writing of the {\it Red Book}. It must be admitted that while very significant progress has been achieved we are still far from a satisfactory understanding of  AGN. Progress is very visible through a number of concrete results, but also from the difference in style and figures between the early 1990s and, for example, the papers presented in these proceedings.

Comparison of the results presented here and elsewhere with the statements included in the {\it Red Book} and discussed in the introduction show that

\begin{itemize}
\item
Source confusion is indeed eliminated. However, weak INTEGRAL sources remain difficult to "identify", showing that  hard X-ray sources weak in terms of INTEGRAL sensitivity are also weak for instruments equipped with imaging  optics at lower energies.
\item
"Hundreds", actually 200, AGN have indeed been observed, albeit at energies $\le $ 100\,keV. It was shown that the flux strongly decreases above this energy for most sources.
\item
Reflection has been observed and is being discussed in terms that differ widely from our expectations at the epoch. Compton processes dominate the emission in the hard X-ray regime.
\item
Hard X-ray observations have proved central to modify and deepen the unification model for Seyfert galaxies.
\item
Although many results exist on the variability of AGN in the hard X-rays, the potential to extract important physical results from the existing and to be acquired data is still very large.
\item
Only very few polarisation results have been obtained with INTEGRAL, none of them for AGN.
\item
Population studies have provided highly relevant results for the understanding of the origin of the extragalactic hard X-ray diffuse background and a significant if small fraction of this background has been resolved in individual sources.

\end{itemize}

INTEGRAL is thus making a very significant contribution to the understanding of AGN physics. This understanding has been progressing more slowly in the last couple of decades than in the years before that. This is mainly due to the fact that the physics of these objects is much more complex than was anticipated. 

The next steps in the study of hard X-ray emission from AGN will come from focusing optics instruments that are now possible in the 10s of keV range. NuStar is the first such instrument, ASTRO-H is expected to be the next one. 

It is an immense privilege to have been associated to the INTEGRAL mission and to have been able to contribute to the progress of our understanding of objects as fascinating as AGN over the past decade.

\end{document}